\documentclass[a4paper,12pt]{article}
\usepackage{epsfig,float}
\usepackage[sort&compress,square,comma]{natbib}
\usepackage[left=2cm,right=2cm]{geometry}
\usepackage{a4wide,graphicx}
\newcommand\T{\rule{0pt}{3ex}}
\newcommand\B{\rule[-1.5ex]{0pt}{0pt}}

\newcommand{\eps}{\epsilon}

\begin{document}

\title{Decay of vector-vector resonances into $\gamma$ and a pseudoscalar meson}

\author{R. Molina$^1$, H. Nagahiro$^2$, A. Hosaka$^3$ and E. Oset$^1$ \\
{\small{\it $^1$Departamento de F\'{\i}sica Te\'orica and IFIC,
Centro Mixto Universidad de Valencia-CSIC,}}\\
{\small{\it Institutos de
Investigaci\'on de Paterna, Apartado 22085, 46071 Valencia, Spain}}\\
{\small{\it $^2$Department of Physics, Nara Women's University, Nara 630-8506, Japan}}\\
{\small{\it $^3$Research Center for Nuclear Physics (RCNP), Osaka University, Ibaraki, Osaka 567-0047, Japan}}\\}

\date{}

\maketitle

 \begin{abstract}
We study the decay of dynamically generated resonances from the interaction of two vectors into a $\gamma$ and a pseudoscalar meson. The dynamics requires anomalous terms involving vertices with two vectors and a pseudoscalar, which renders it special. We compare our result with data on $K^{*+}(1430)\to K^+\gamma$ and $K^{*0}(1430)\to K^0\gamma$ and find a good agreement with the data for the  $K^{*+}(1430)$ case and a width considerably smaller than the upper bound measured for the  $K^{*0}(1430)$ meson.
\end{abstract}

\section{Introduction}

The radiative decay of a mesonic state has long been argued to be
crucial in the determination of the nature of the
state~\cite{Pennington:2008qa}. For instance, the non-observation of
the $f_0(1500)$ decaying into two photons has been used to support
its dominant glue nature~\cite{Amsler:2002ey}. In this sense the radiative decay has also been advocated as a tool to determine the molecular nature of many mesonic and baryonic states \cite{Doring:2007rz,Branz:2008ha,Dong:2009yp,Branz:2010sh,Dong:2010xv,Branz:2010gd,liuke}.

    The dynamical generation of many mesonic states using chiral unitary dynamics \cite{review,puri} has stimulated further studies of these decay modes of mesonic states. Though scalar mesons have been for long in the list of dynamically generated mesons from the interaction of two pseudoscalar mesons 
\cite{npa,Kaiser:1998fi,Oller:1997ng,ramonet,Markushin:2000fa}, it has been only very recently that the systems of two vector mesons have been investigated
\cite{raquel,geng,raquelnaga,xyz,taniacharm}, and many dynamically generated states have appeared which have been associated with known resonances of the PDG \cite{pdg}. In this sense the $f_0(1370)$, $f_2(1270)$, $f'_2(1525)$, $f_0(1710)$ and $K^*_2(1430)$ were generated in \cite{raquel,geng} from the interaction of the nonet of vector mesons with itself. Similarly, some D* mesons are generated in \cite{raquelnaga}, the $D^*_{s2}(2573)$ among other states is generated in \cite{taniacharm} and some hidden charm states, some of which could be identified with the new X,Y,Z resonances recently reported, were also found in \cite{xyz}. The task of studying several radiative decay modes is important for these states in view of the fact that many of them have traditionally been accommodated within quark models without much difficulty  
\cite{Li:2000zb,klempt,crede,isgur,Barnes:1996ff,Barnes:2002mu,Anisovich:2002im}.  Support for the new nature of these states comes gradually from studies of different decay modes. In this sense in \cite{yamagata} the radiative decay of the $f_0(1370)$ and $f_2(1270)$ resonances into $\gamma \gamma$, was studied and good results compared with experiment were found. These studies were extended to the SU(3) states, and also good agreement with experiment was found in the cases where there was available data for comparison \cite{Branz:2009cv}. The study of the $J/\psi \to \phi (\omega) f_2(1270)$, $J/\psi \to \phi
  (\omega) f'_2(1525)$ and $J/\psi \to K^{*0}(892) \bar{K}^{* 0}_2(1430)$ decay in \cite{jpsiplb}, and $J/\psi$ decay into  $\gamma f_2(1270)$, $\gamma f_2'(1525)$, $\gamma f_0(1370)$ and $\gamma f_0(1710)$ in \cite{Geng:2009iw} has given extra support to the claimed nature of these resonances as vector-vector bound states or resonances.
  
     In this paper  we pose a new challenge to this idea by investigating the decay of the dynamically generated states of \cite{geng} into a pseudoscalar meson and one baryon. As we shall see, the decay proceeds via anomalous interaction terms which involve two vectors and a pseudoscalar meson, and thus one is exploring dynamics quite different than the one needed in other alternative studies, like in quark models. 
     
     Although there are not many data to compare, we shall see that the agreement with them is satisfactory, within theoretical and experimental uncertainties, and the results obtained should encourage further theoretical and experimental studies in other sectors, like charm or hidden charm mesons.

\section{Formalism}
In this work we study the radiative decay of the VV dynamically generated resonances found in \cite{geng} into P$\gamma$. In Table \ref{tab:geng}, we display the masses and widths obtained in that work and the experimental counterpart of each resonance in the assignment made in \cite{geng}. We see in this table that eleven resonances were found, five of them were associated with resonances that appear in the PDG \cite{pdg}. First of all, we consider all the possible cases of spin-parity of the initial meson in Table \ref{tab:geng}: In case we had an initial meson with $J^P=0^+$, the angular momentum between the pseudoscalar meson and photon should be $L=1$, which implies negative final parity, and is not allowed. In the language of photon multipoles this corresponds to an M0 transition, which does not exist. The rest of the resonances in Table \ref{tab:geng} are either with or without strangeness. The ones without strangeness except for those of $J=1$ have positive C-parity which does not allow the decay into $P\gamma$. This leaves non vanishing decay rates only for the $h_1$, $b_1$, $K_1$ and $K^*_2(1430)$, with only this latter one having a clear experimental counterpart, the $K^*_2(1430)$. In the present work we concentrate on this latter case, where there are also experimental data in the PDG for its decay into P$\gamma$: 
\begin{eqnarray}
\Gamma(K^{*+}_2(1430)\to K^+\gamma)/\Gamma&=&(2.4\pm 0.5)\times 10^{-3}\nonumber\\
\Gamma(K^{*0}_2(1430)\to K^0\gamma)/\Gamma&<&9\times 10^{-4}\label{eq:widex}
\end{eqnarray}
\begin{table*}[htpb]
      \renewcommand{\arraystretch}{1.5}
     \setlength{\tabcolsep}{0.1cm}
\begin{center}
\begin{tabular}{c|c|cc|ccc}
$I^{G}(J^{PC})$&\multicolumn{3}{c|}{Theory} & \multicolumn{3}{c}{PDG data}\\\hline\hline
              & Pole position &\multicolumn{2}{c|}{Real axis} & Name & Mass & Width  \\
              &               & $\Lambda_b=1.4$ GeV & $\Lambda_b=1.5$ GeV &           \\\hline
$0^+(0^{++})$ & (1512,51) & (1523,257) & (1517,396)& $f_0(1370)$ & 1200$\sim$1500 & 200$\sim$500\\
$0^+(0^{++})$ & (1726,28) & (1721,133) & (1717,151)& $f_0(1710)$ & $1724\pm7$ & $137\pm 8$\\
$0^-(1^{+-})$ & (1802,78) & \multicolumn{2}{c|} {(1802,49)}   & $h_1$\\
$0^+(2^{++})$ & (1275,2) & (1276,97) & (1275,111) & $f_2(1270)$ & $1275.1\pm1.2$ & $185.0^{+2.9}_{-2.4}$\\
$0^+(2^{++})$ & (1525,6) & (1525,45) &(1525,51) &$f_2'(1525)$ & $1525\pm5$ & $73^{+6}_{-5}$\\\hline
$1^-(0^{++})$    & (1780,133) & (1777,148) &(1777,172) & $a_0$\\
$1^+(1^{+-})$    & (1679,235) & \multicolumn{2}{c|}{(1703,188)} & $b_1$ \\
$1^-(2^{++})$    &  (1569,32) & (1567,47) & (1566,51)& $a_2(1700)??$
\\\hline
$1/2(0^+)$       &  (1643,47) & (1639,139) &(1637,162)&  $K_0^*$ \\
$1/2(1^+)$       & (1737,165) &  \multicolumn{2}{c|}{(1743,126)} & $K_1(1650)?$\\
$1/2(2^+)$       &  (1431,1) &(1431,56) & (1431,63) &$K_2^*(1430)$ & $1429\pm 1.4$ & $104\pm4$\\
 \hline\hline
    \end{tabular}
\end{center}
\caption{The properties, (mass, width) [in units of MeV], of the 11  dynamically
generated states and, if existing, of those of their PDG
counterparts. Theoretical masses and widths are obtained from two
different ways: ``pole position'' denotes the numbers obtained from
the pole position on the complex plane, where the mass corresponds to the
real part of the pole position and the width corresponds to twice
the imaginary part of the pole position (the box diagrams
corresponding to decays into two pseudoscalars are not included);
"real axis" denotes the results obtained from the real axis amplitudes
squared, where the mass corresponds to the energy at which the amplitude
squared has a maximum and the width corresponds to the difference
between the two energies, where the amplitude squared is half of the
maximum value.}
\label{tab:geng}
\end{table*}
From \cite{geng} we take the channels and coupling constants, $g_i$, of the $K^*_2(1430)$, that are shown in Table \ref{tab:geng2}. As we see in this table, the $K^*_2(1430)$ couples to three channels: $\rho K^*$,  $K^*\omega$ and $K^*\phi$, the coupling to $\rho K^*$ being considerably larger than for the other two channels.
\begin{table*}[htpb]
      \renewcommand{\arraystretch}{1.5}
     \setlength{\tabcolsep}{0.1cm}
     \centering
     \begin{tabular}{c|ccc}
  $\sqrt{s}_{\mathrm{pole}}$&\multicolumn{3}{c}{$g_i$  [spin $=2$]}\\\hline\hline
       &  $ K^* \rho$ & $K^*\omega$ & $K^*\phi$    \\\hline
 $(1431,-i1)$ &  $(10901,-i71)$ & $(2267,-i13)$ & $(-2898,i17)$  \\
 \hline\hline
    \end{tabular} 
    \caption{Pole positions and residues in the strangeness $= 0$ and isospin $=0$ channel.
             All the quantities are in units of MeV.}\label{tab:geng2}
       \end{table*}
       \begin{figure}[ht]
       \begin{center}
\includegraphics[scale=0.7]{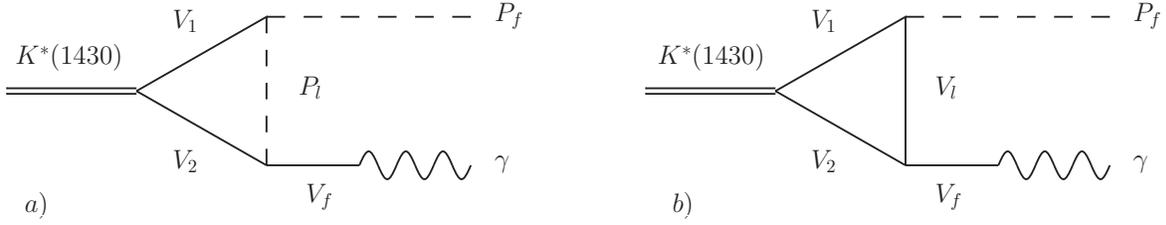} 
\end{center}
\caption{The two different diagrams that contribute to the $K^*(1430)\to K\gamma$ decay.}
\label{fig:diag}
\end{figure}
       
\begin{figure}[ht]
       \begin{center}
\includegraphics[scale=0.7]{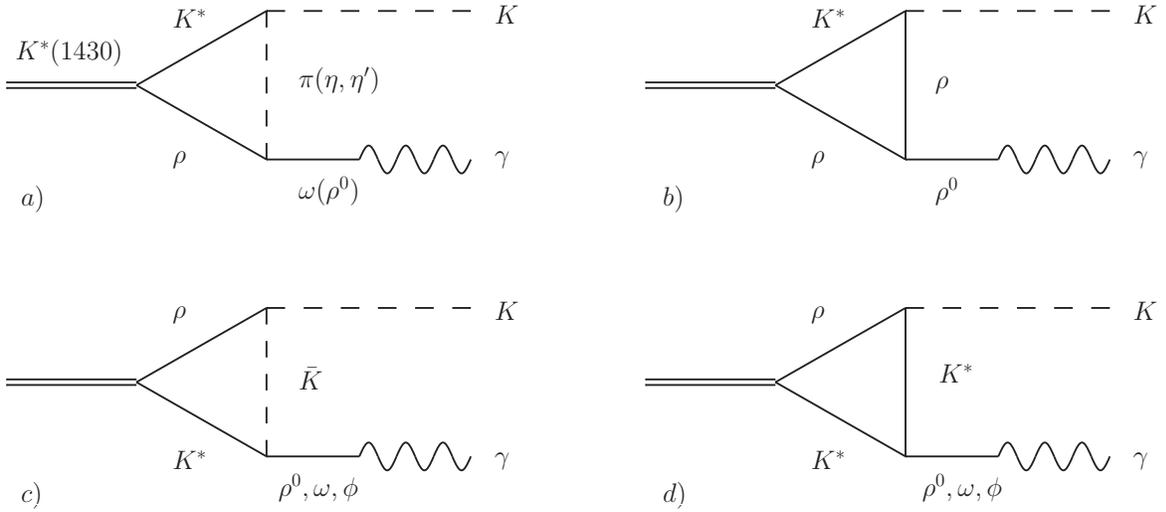} 
\end{center}
\caption{Possible Feynman diagrams in the isospin basis contributing to the $K^*(1430)\to K\gamma$ decay in the $\rho K^*$ channel.}
\label{fig:diag2}
\end{figure}

In Fig. \ref{fig:diag} we show the two kinds of Feynman diagrams that lead to the decay of a resonance into P$\gamma$ in the VV molecular picture that combines HGS (Hidden local Gauge Symmetry) \cite{hidden1,hidden2,hidden3,hidden4,roca} and unitarity \cite{raquel,geng}. The two different diagrams contain an anomalous VVP coupling, whereas they can be distinguished from the exchange of one pseudoscalar meson, $P_l$, containing a PPV vertex, see Fig. \ref{fig:diag} a), or a vector meson, $V_l$, with a 3V vertex, as shown in Fig. \ref{fig:diag} b). These two kinds of diagrams lead to four possible configurations, as shown in Fig. \ref{fig:diag2} for the $\rho K^*$ channel, depending on whether $P_l(V_l)$ is a non-strange meson, Fig. \ref{fig:diag2} a) and b), or a strange meson, Fig. \ref{fig:diag2} c) and d). At the end, all possible $VV$ channels are taken into account.

The V$\gamma$, PPV and 3V vertices are provided by the hidden gauge formalism, where a photon always comes out from a vector meson. In the HGS formalism, the vector meson fields are
gauge bosons of a hidden local symmetry transforming inhomogeneously and chiral symmetry is preserved \cite{hidden1,hidden2,hidden3,hidden4,roca}. The HGS Lagrangian involving pseudoscalar, vector mesons and photons is
\begin{equation}
{\cal L}={\cal L}^{(2)}+ {\cal L}_{III}
\end{equation}
with
\begin{eqnarray}
 {\cal L}^{(2)}&=&\frac{1}{4}f^2\langle D_\mu U D^\mu U^\dagger+
 \chi U^\dagger+\chi^\dagger U\rangle \\
 {\cal L}_{III}&=&-\frac{1}{4}\langle V_{\mu\nu}V^{\mu\nu}\rangle
 +\frac{1}{2}M_V^2\langle [V_\mu-\frac{i}{g}\Gamma_\mu]^2\rangle , 
 \label{eq:laghg}
 \end{eqnarray}
where $\langle ...\rangle$ represents a trace over SU(3) matrices. The
covariant derivative is defined by
\begin{equation}
D_\mu U=\partial_\mu U-ieQA_\mu U+ieUQA_\mu,
\end{equation}
with $Q=diag(2,-1,-1)/3$, $e=-|e|$ the electron charge, and $A_\mu$
the photon field.
The chiral matrix $U$ is given by
\begin{equation}
U=e^{i\sqrt{2}P/f}\ ,
\end{equation}
where the P matrix contains
the nonet of the pseudoscalars in the physical basis considering $\eta$, $\eta'$ mixing \cite{gamphi3770}:
\begin{equation}
\renewcommand{\tabcolsep}{1cm}
\renewcommand{\arraystretch}{2}
P=\left(
\begin{array}{ccc}
\frac{\eta}{\sqrt{3}}+\frac{\eta'}{\sqrt{6}}+\frac{\pi^0}{\sqrt{2}} & \pi^+ & K^+\\
\pi^- &\frac{\eta}{\sqrt{3}}+\frac{\eta'}{\sqrt{6}}-\frac{\pi^0}{\sqrt{2}} & K^{0}\\
K^{-} & \bar{K}^{0} &-\frac{\eta}{\sqrt{3}}+\sqrt{\frac{2}{3}}\eta'
\end{array}
\right)\ ,
\end{equation}
and $V_\mu$ represents the vector nonet:
 \begin{equation}
\renewcommand{\tabcolsep}{1cm}
\renewcommand{\arraystretch}{2}
V_\mu=\left(
\begin{array}{ccc}
\frac{\omega+\rho^0}{\sqrt{2}} & \rho^+ & K^{*+}\\
\rho^- &\frac{\omega-\rho^0}{\sqrt{2}} & K^{*0}\\
K^{*-} & \bar{K}^{*0} &\phi
\end{array}
\right)_\mu\ .
\end{equation}
In  ${\cal L}_{III}$, $V_{\mu\nu}$ is defined as 
\begin{equation} V_{\mu\nu}=\partial_\mu
V_\nu-\partial_\nu V_\mu-ig[V_\mu,V_\nu]
\end{equation}
 and 
\begin{equation}
\Gamma_\mu=\frac{1}{2}[u^\dagger(\partial_\mu-ieQA_\mu)u
+u(\partial_\mu-ieQA_\mu)u^\dagger]
\end{equation}
with $u^2=U$. 
The value of the coupling constant $g$ of the Lagrangian Eq.~(\ref{eq:laghg})
is
\begin{equation}
g=\frac{M_V}{2f},
\end{equation}
with $M_V$ the vector meson mass and $f=93$ MeV the pion decay constant. Other properties of $g$ are \cite{Ecker,roca}:
\begin{equation}
 \frac{F_V}{M_V}=\frac{1}{\sqrt{2}g}\qquad,\qquad
\frac{G_V}{M_V}=\frac{1}{2\sqrt{2}g}\qquad,\qquad
F_V=\sqrt{2}f\qquad,\qquad G_V=\frac{f}{\sqrt{2}}\ .
\label{eq:VMDrelations}
\end{equation}
Eq. (\ref{eq:laghg}) provides the following terms:
\begin{eqnarray}
\mathcal{L}_{V\gamma}&=&-M_V^2\frac{e}{g}A_\mu\langle V^\mu Q\rangle\nonumber\\
\mathcal{L}_{PPV}&=&-ig\langle V^\mu [P,\partial_\mu P]\rangle
\label{lag1}
\end{eqnarray}
and
\begin{eqnarray}
\mathcal{L}_{3V}=
&=&ig\langle (V^\mu \partial_\nu V_\mu -\partial_\nu V_\mu V^\mu)
V^\nu)\rangle
\label{lag2}
\end{eqnarray}

Both diagrams in Fig. \ref{fig:diag} contain an anomalous VVP vertex, which in principle one
could expect to be small due to the higher order nature of the anomalous term in the chiral
expansion. This anomalous VVP interaction accounts for a process that does
not preserve intrinsic parity, and can be obtained from the gauged Wess-Zumino term (see
e.g. \cite{Meissner,Pallante}). However, as the relevant energy becomes larger, the role of the anomalous contribution
becomes more important as it contains momentum factors (see Eq.~(\ref{lag3})).
This has also been
seen in works on the
radiative decays of scalar mesons \cite{Nagahiro,Branz}. The VVP Lagrangian is \cite{Pallante,Bramon,Pelaez}:
\begin{equation}
\mathcal{L}_{VVP}=\frac{G'}{\sqrt{2}}\epsilon^{\mu\nu\alpha\beta}\langle\partial_\mu V_\nu \partial_\alpha V_\alpha P\rangle
\label{lag3}
\end{equation}
with $G' = 3g'^2/(4\pi^2f)$ and $g' = -G_V M_\rho/(\sqrt{2}f^2)$. In the following subsections we evaluate the two different kind of diagrams shown in Fig. \ref{fig:diag}.
\subsection{Diagram of the $K^*(1430)\to K\gamma$ decay containing the $PPV$ vertex}

\begin{figure}[ht]
       \begin{center}
\includegraphics[scale=0.7]{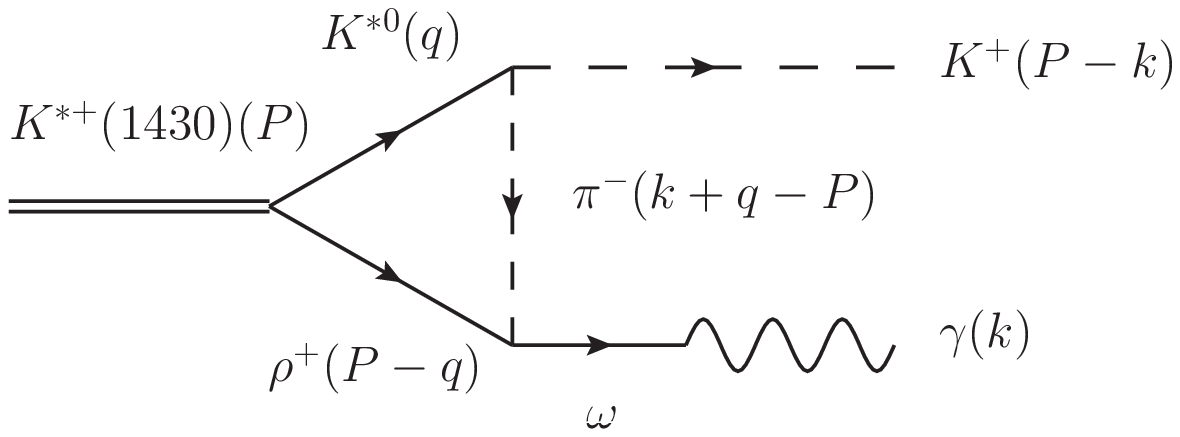} 
\end{center}
\caption{Feynman diagram of the $K^{*+}(1430)\to K^+\gamma$ decay in the $\rho^+ K^{*0}$ channel with a $PPV$ vertex.}
\label{fig:ppv}
\end{figure}

\begin{figure}[ht]
       \begin{center}
\includegraphics[scale=0.7]{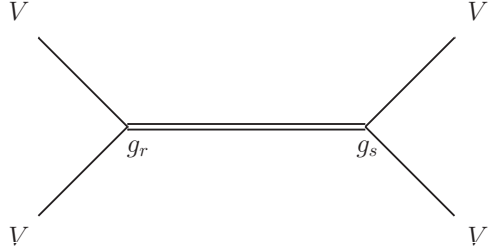} 
\end{center}
\caption{Dynamically generated resonance from the $VV$ interaction}
\label{fig:new}
\end{figure}

In Fig. \ref{fig:ppv} we show the first diagram to compute in charge basis with explicit momentum. In what follows, we shall consider the $K^{*+}(1430)$ at rest. First of all, we need the coupling of the resonance $K^{*+}(1430)$ to $K^{*0}\rho^+$. This coupling is given by the approach of \cite{geng}, where the coupling is calculated as the residue of the $VV\to VV$ amplitude in the pole position of the resonance (see Fig. \ref{fig:new}), which close to a pole can be expressed as:
\begin{eqnarray}
t^{(J=2)ij}_{rs}&=&\frac{g_r g_s}{s-s_{\mathrm{pole}}}\,\lbrace \frac{1}{2}(\epsilon^{(1)i}\epsilon^{(2)j}+\epsilon^{(1)j}\epsilon^{(2)i})-\frac{1}{3}\epsilon^{(1)l}\epsilon^{(2)l}\delta^{ij}\rbrace\,\nonumber\\&\times&\lbrace \frac{1}{2}(\epsilon^{(1)i}\epsilon^{(2)j}+\epsilon^{(1)j}\epsilon^{(2)i})-\frac{1}{3}\epsilon^{(1)m}\epsilon^{(2)m}\delta^{ij}\rbrace\ ,\label{lag4}
\end{eqnarray}
with $s_{\mathrm{pole}}=(M-i\Gamma/2)^2$. We can see in this amplitude, by looking at the diagram in Fig. \ref{fig:new}, that the coupling of the resonance to a $VV$ channel is given by 
$\tilde{g_r}=
g_r\lbrace\frac{1}{2}(\epsilon^{(1)i}\epsilon^{(2)j}+\epsilon^{(1)j}\epsilon^{(2)i})-\frac{1}{3}\epsilon^{(1)l}\epsilon^{(2)l}\delta^{ij}\rbrace$,
$r$ or $s$ corresponding to one of the channels $\rho K^*$, $\omega K^*$ or $\phi K^*$. These couplings are given in Table \ref{tab:geng2} in the isospin basis and we have to multiply them for the correspondent Clebsch Gordan coefficient:
\begin{eqnarray}
\vert\rho\,K^*,1/2,1/2\rangle&=&-\sqrt{\frac{2}{3}}\,\rho^+ K^{*0}-\frac{1}{\sqrt{3}}\,\rho^0 K^{*+}\nonumber\\
\vert\rho\,K^*,1/2,-1/2\rangle&=&-\sqrt{\frac{2}{3}}\,\rho^- K^{*+}+\frac{1}{\sqrt{3}}\,\rho^0 K^{*0}\ .\nonumber\\
\end{eqnarray}
The isospin coefficient is denoted as $g_I$. Thus, from Eqs. (\ref{lag1}), (\ref{lag2}), (\ref{lag3}) and (\ref{lag4}) we can write the vertices involved in the diagram of Fig. \ref{fig:ppv} as
\begin{eqnarray}
t_{RV_1V_2}^{ij}&=&g_Ig_r\lbrace\frac{1}{2}(\epsilon^{(1)i}\epsilon^{(2)j}+\epsilon^{(1)j}\epsilon^{(2)i})-\frac{1}{3}\epsilon^{(1)l}\epsilon^{(2)l}\delta^{ij}\rbrace\nonumber\\
t_{V_f\gamma}&=&\lambda \,\frac{e}{g}M_{V_f}^2\epsilon_\mu^{(\gamma)}\epsilon^{(f)\mu}\nonumber\\
t_{P_lP_fV_1}&=&Ag(p_{\mathrm{in}}+p_{\mathrm{fin}})_\mu\eps^{(1)\mu}=-Ag(2(P-k)-q)_\mu\eps^{(1)\mu}\nonumber\\
t_{V_2V_fP_l}&=&-B\frac{G'}{\sqrt{2}}\eps^{\alpha\beta\gamma\delta}(P-q)_\alpha\eps_{\beta}^{(2)}k_\gamma\eps_\delta^{(f)}\label{eq:ts}\ ,
\end{eqnarray}
with $V_1=K^{*0}$, $V_2=\rho^+$, $V_f=\omega$, $P_l=\pi^-$, $P_f=K^+$, and the coefficients $g_I$, $g_r$, $A$, $B$ and $\lambda$ are: $-\sqrt{\frac{2}{3}}$, $(10901,-i71)$ MeV, $-1$, $\sqrt{2}$ and $\frac{1}{3\sqrt{2}}$ respectively. The $V_f\to\gamma$ conversion essentially replaces, up to a constant, $\eps_\delta^{(f)}$ by $\eps_\delta^{(\gamma)}$. Therefore, we can write the amplitude of the diagram depicted in Fig. \ref{fig:ppv} as
\begin{eqnarray}
-it_{K^{*+}(1430)\to K^+\gamma}^{ij}&=&\int \frac{d^4 q}{(2\pi)^4}\lbrace\frac{1}{2}(\epsilon^{(1)i}\epsilon^{(2)j}+\epsilon^{(1)j}\epsilon^{(2)i})-\frac{1}{3}\epsilon^{(1)}_l\epsilon^{(2)}_l\delta^{ij}\rbrace \nonumber\\&\times &\eps^{(1)\mu}(2(P-k)-q)_\mu \eps^{\alpha\beta\gamma\delta}(P-q)_\alpha\eps_\beta^{(2)}k_\gamma\eps_\delta^{(\gamma)}\nonumber\\&\times & \frac{1}{q^2-M_{1}^2+i\eps}\frac{1}{(k+q-P)^2-m_l^2+i\eps}\nonumber\\&\times&\frac{1}{(P-q)^2-M_2^2+i\eps}\times\mathrm{F_I}\times e g_r G'\ ,
\end{eqnarray}
with $M_{1}=m_{K^*}$, $M_2=m_\rho$, $m_l=m_\pi$ and
$\mathrm{F_I}=\frac{1}{\sqrt{2}}AB\lambda g_I=\frac{1}{3\sqrt{3}}$. We
should be consistent with the approximation done in \cite{geng,raquel},
where $\vert\vec{q}\,\vert/M_1\simeq 0$, which implies that
$\eps^{(1)0}\simeq 0$. This means that the $\mu$ and $\beta$ indices
should be spatial and also that the $q^i q^j/M_V^2$ terms in the sum
over vector polarizations should be neglected. For convenience, we will
keep them as covariant indices and will consider them 
as spatial indices
at the end. Thus, after summing over polarizations
\begin{eqnarray}
\sum_\lambda\eps^{(1)i}\eps^{(1)\mu}=-g^{i\mu}\nonumber\\
\sum_\lambda\eps^{(2)j}\eps_{\beta}^{(2)}=-g^{j}_{\hspace{0.15cm} \beta}\ ,
\end{eqnarray}
we get
\begin{eqnarray}
-it_{K^{*+}(1430)\to K^+\gamma}^{ij}&=&\int \frac{d^4 q}{(2\pi)^4}\lbrace  \frac{1}{2}\eps^{\alpha j\gamma\delta}(2(P-k)-q)^i(P-q)_\alpha k_\gamma \eps_\delta^{(\gamma)}\nonumber\\&+&\frac{1}{2}\eps^{\alpha i\gamma\delta}(2(P-k)-q)^j(P-q)_\alpha k_\gamma \eps_\delta^{(\gamma)}\nonumber\\&-&\frac{1}{3}\eps^{\alpha m\gamma\delta}(2(P-k)-q)^m(P-q)_\alpha k_\gamma \eps_\delta^{(\gamma)}\delta^{ij}\rbrace \nonumber\\&\times &\frac{1}{q^2-M_{1}^2+i\eps}\frac{1}{(k+q-P)^2-m_l^2+i\eps}\nonumber\\&\times &\frac{1}{(P-q)^2-M_2^2+i\eps}\times\mathrm{F_I}\times e g_r G'\ .\label{eq:three}
\end{eqnarray}
All the terms of Eq. (\ref{eq:three}) are proportional to an integral like
\begin{eqnarray}
& &\int \frac{d^4 q}{(2\pi)^4}(2(P-k)-q)^i(P-q)_\alpha \frac{1}{q^2-M_{1}^2+i\eps}\nonumber\\&\times &\frac{1}{(k+q-P)^2-m_l^2+i\eps}\frac{1}{(P-q)^2-M_2^2+i\eps}\ ,\label{eq:int}
\end{eqnarray} 
which from Lorentz covariance must be a tensor built from $P$ and $k$,
\begin{eqnarray}
a g^i_{\hspace{0.15cm}\alpha}+b P^iP_\alpha +ck^i P_\alpha+d P^i k_\alpha +e k^i k_\alpha\label{eq:lor}\ .
\end{eqnarray}
The second and fourth terms in Eq. (\ref{eq:lor}) vanish directly because $P^i=0$. After substituting the integrals of Eq. (\ref{eq:three}) by Eq. (\ref{eq:lor}) with the correspondent indices, we see that the first term in Eq. (\ref{eq:lor}) leads to a term proportional to
\begin{eqnarray}
\frac{1}{2}k_\gamma\eps_{\delta}^{(\gamma)}a(\eps^{ij\gamma\delta}+\eps^{ji\gamma\delta})-\frac{1}{3}\eps^{\alpha m \gamma\delta}k_\gamma \eps_\delta^{(\gamma)}\delta^{ij}a g^m_{\hspace{0.15cm}\alpha}\label{eq:1}\ .
\end{eqnarray}
This term vanishes when one contracts the antisymmetric operator
$\eps^{\alpha m\gamma\delta}$ with the symmetric
$g^m_{\hspace{0.15cm}\alpha}$. This is a welcome feature because this
term of the integral in Eq. (\ref{eq:int}) was the only one that is divergent. The fifth term, $e k^i k_\alpha$, leads to terms proportional to $k_\gamma k_\alpha \eps^{\alpha l \gamma \delta}$, and therefore it also vanishes. The third term, $ck^i P_\alpha$, is the only one that remains, but we can still simplify it a little bit. The integral in Eq. (\ref{eq:three}) is proportional to
\begin{eqnarray}
\frac{1}{2}c P_\alpha k_\gamma\eps_\delta^{(\gamma)}(k^i\eps^{\alpha j\gamma \delta}+k^j\eps^{\alpha i\gamma \delta})-\frac{1}{3}c\eps^{\alpha m\gamma\delta}k_\gamma\delta^{ij}\eps_\delta^{(\gamma)}k^m P_\alpha\ .
\label{eq:c1}
\end{eqnarray}
The last term in the above equation vanishes for $P^i=0$. To see it, let us split the factor $\eps^{\alpha m\gamma \delta} k_\gamma k^m P_\alpha$ in two terms
\begin{eqnarray}
\sum_{m=1,3}\eps^{\alpha m 0\delta}k_0 k^m P_\alpha+\sum_{m=1,3}\sum_{l=1,3}\eps^{\alpha m l\delta}k_l k^m P_\alpha\ ,
\end{eqnarray} 
the last term is zero since it is a product of an antisymmetric operator with a symmetric one. In addition, the presence of $P_\alpha$ forces $\alpha=0$, which makes the first term also disappear.

Now we must evaluate the $c$ coefficient. Let us use the formula of the Feynman parametrization for $n=3$
\begin{equation}
\frac{1}{\alpha\beta\gamma}=2\int^1_0 dx\int^x_0 dy\frac{1}{[\alpha+(\beta-\alpha)x+(\gamma-\beta)y]^3}\ .\label{eq:fein}
\end{equation}
For the integral of Eq. (\ref{eq:int}), we can use the above parametrization with 
\begin{eqnarray}
\alpha&=&q^2-M^2_1\nonumber\\
\beta&=&(P-q)^2-M^2_2\nonumber\\
\gamma&=&(P-q-k)^2-m_l^2\label{eq:abc}\ .
\end{eqnarray}
Besides this, we define a new variable $q'=q-Px+ky$, such that the integral of Eq. (\ref{eq:int}) can be expressed as
\begin{eqnarray}
2\int \frac{d^4 q'}{(2\pi)^4} \int^1_0 dx\int^x_0 dy  (2(P-k)-q)^i(P-q)_\alpha \frac{1}{(q'^2+s)^3}\ ,\label{eq:int2}
\end{eqnarray} 
with 
\begin{eqnarray}
s=-(P^0)^2 x^2+2P^0k^0 xy+((P^0)^2-M_2^2+M^2_1)x+(-2P^0k^0+M^2_2-m_l^2)y-M^2_1\label{eq:s}\ .
\end{eqnarray}
>From Eq. (\ref{eq:int2}), we must take the $k^i P_\alpha$ term. Therefore,
\begin{eqnarray}
c=2\int \frac{d^4 q'}{(2\pi)^4} \int^1_0 dx\int^x_0 dy \frac{(1-x)(y-2)}{(q'^2+s)^3}\ .\label{eq:c2}
\end{eqnarray}
Now we still can perform the integral in the $q'$ variable analytically:
\begin{equation}
\int d^4 q'\frac{1}{(q'^2+s)^3}=\frac{i\pi^2}{2s}\label{eq:int31}\ ,
\end{equation}
and finally, we get
\begin{equation}
c=\frac{i}{16\pi^2} \int^1_0 dx\int^x_0 dy \frac{(1-x)(y-2)}{s}\label{eq:c}\ ,
\end{equation}
and the amplitude of the diagram of Fig. \ref{fig:ppv} as
\begin{eqnarray}
-it_{K^{*+}(1430)\to K^+\gamma}^{ij}&=&\frac{1}{2}c P_\alpha k_\gamma\eps_\delta^{(\gamma)}(k^i\eps^{\alpha j\gamma \delta}+k^j\eps^{\alpha i\gamma \delta})\mathrm{F_I}\,e g_r G'\label{eq:dia1}\ .
\end{eqnarray}
\subsection{Diagram of the $K^*(1430)\to K\gamma$ decay containing the $3V$ vertex}
\begin{figure}[ht]
       \begin{center}
\includegraphics[scale=0.7]{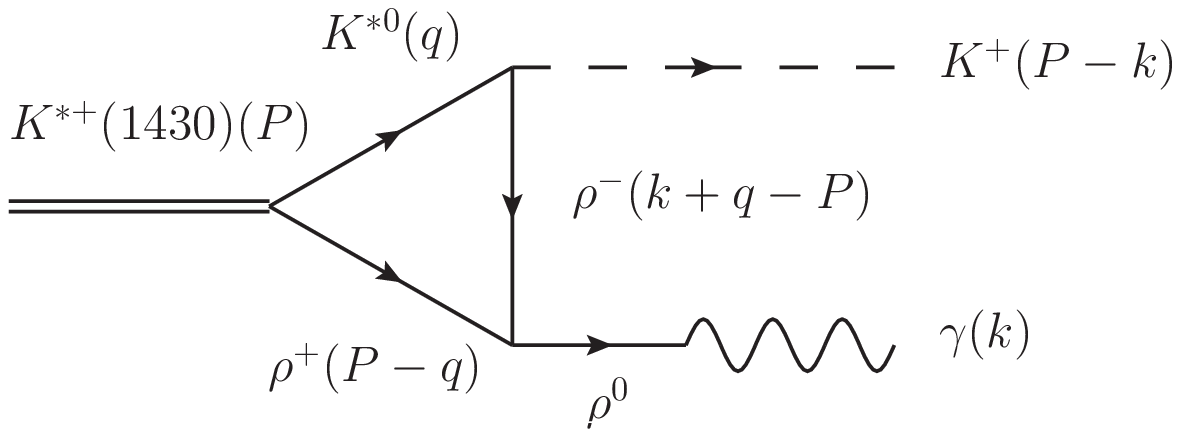} 
\end{center}
\caption{Feynman diagram of the $K^{*+}(1430)\to K^+\gamma$ decay in the $\rho^+ K^{*0}$ channel with a $3V$ vertex.}
\label{fig:3v}
\end{figure}

Now, we want to compute the second diagram for the $K^*(1430)\to K\gamma$ decay depicted in Fig. \ref{fig:diag2}. In Fig. \ref{fig:3v} we show this diagram with the explicit momenta in the case of the $K^{*0}\rho^+$ intermediate state. The difference with the diagram calculated in the previous section is the presence of the $3V$ vertex, which we can obtain from the Lagrangian of Eq. (\ref{lag2}). This vertex and the anomalous $VVP$ vertex are 
\begin{eqnarray}
t_{V_2V_lV_f}&=&gD\lbrace(2k+q-P)_\mu\eps_\nu^{(l)}\eps^{(2)\mu}\eps^{(f)\nu}\nonumber\\
&-&(k+P-q)_\mu\eps_\nu^{(2)}\eps^{(l)\mu}\eps^{(f)\nu}\nonumber\\
&+&(2(P-q)-k)_\mu\eps_\nu^{(l)}\eps^{(f)\mu}\eps^{(2)\nu}\rbrace\nonumber\\
t_{V_1V_lP_f}&=&-B\frac{G'}{\sqrt{2}}\eps^{\alpha\beta\gamma\delta}q_\alpha\eps_{\beta}^{(1)}(k+q-P)_\gamma\eps_\delta^{(l)}\ ,
\end{eqnarray}
 with $D=\sqrt{2}$, $B=1$, and $g_I$, $g_r$, $\lambda$ in Eqs. (\ref{eq:ts}) are $-\sqrt{\frac{2}{3}}$, $(10901,-i71)$ MeV, $\frac{1}{\sqrt{2}}$ respectively. With this, we can write the amplitude of the diagram in Fig. \ref{fig:3v} as
\begin{eqnarray}
-it_{K^{*+}(1430)\to K^+\gamma}^{ij}&=&\int \frac{d^4q}{(2\pi)^4}\lbrace\frac{1}{2}(\epsilon^{(1)i}\epsilon^{(2)j}+\epsilon^{(1)j}\epsilon^{(2)i})-\frac{1}{3}\epsilon^{(1)}_l\epsilon^{(2)}_l\delta^{ij}\rbrace\nonumber\\
&\times &\eps^{\alpha\beta\gamma\delta} q_\alpha\eps_\beta^{(1)}(k+q-P)_\gamma\eps_\delta^{(l)}\nonumber\\
&\times &\lbrace(2k+q-P)_\mu\eps_\nu^{(l)}\eps^{(2)\mu}\eps^{(\gamma)\nu}-(k+P-q)_\mu\eps_\nu^{(2)}\eps^{(l)\mu}\eps^{(\gamma)\nu}\nonumber\\&+&(2(P-q)-k)_\mu\eps_\nu^{(l)}\eps^{(\gamma)\mu}\eps^{(2)\nu}\rbrace\nonumber\\&\times &\frac{1}{q^2-M^2_1+i\eps}\frac{1}{(k+q-P)^2-M^2_l+i\eps}\nonumber\\&\times &\frac{1}{(P-q)^2-M^2_2+i\eps}\times \mathrm{F_I'}\times eg_r G'
\end{eqnarray}
with $\mathrm{F_I'}=-\frac{1}{\sqrt{2}}g_I B D\lambda$. The way to proceed is very similar to that of the previous subsection, with the only difference in the use of the Lorentz condition, $k_\mu \eps^{(\gamma)\mu}=0$. Now, we get two kinds of integrals. The first one is
\begin{eqnarray}
 \int\frac{d^4q}{(2\pi)^4} q_\alpha \frac{1}{q^2-M^2_1+i\eps}\frac{1}{(k+q-P)^2-M^2_l+i\eps}\frac{1}{(P-q)^2-M^2_2+i\eps}\ , \label{eq:int3}
\end{eqnarray}
which from Lorentz covariance takes the form
\begin{eqnarray}
a_1 P_\alpha+b_1 k_\alpha \ .
\end{eqnarray}
After contracting with the term $k_\gamma P_\delta \eps^{\alpha i \gamma \delta}$, this integral becomes zero. The second integral is 
\begin{eqnarray}
\int\frac{d^4q}{(2\pi)^4} q_\alpha (2k+q-P)^j\frac{1}{q^2-M^2_1+i\eps}\frac{1}{(k+q-P)^2-M^2_l+i\eps}\frac{1}{(P-q)^2-M^2_2+i\eps} 
\label{eq:int4}
\end{eqnarray}
which takes the form
\begin{eqnarray}
a_2 g_{\hspace{0.15cm}\alpha }^j+b_2 k_\alpha k^j+c_2k^j P_\alpha+d_2 P^j k_\alpha+e_2 P^j P_\alpha \ .
\end{eqnarray}
The last two terms are zero since $P^j=0$, and the first one disappears because it gives rise to the factor $g_{\hspace{0.15cm}\alpha}^j\eps^{\alpha i\gamma\delta}+g_{\hspace{0.15cm}\alpha }^i\eps^{\alpha j\gamma\delta}=0$. The final amplitude is a function of the $b_2$ and $c_2$ coefficients, and it can be expressed as
\begin{eqnarray}
-it_{K^{*+}(1430)\to K^+\gamma}^{ij}&=&-\frac{1}{2}(k^j\eps^{\alpha i\gamma\delta}+k^i\eps^{\alpha j\gamma\delta})\eps_\delta^{(\gamma)}(-b_2k_\alpha P_\gamma+c_2P_\alpha k_\gamma)\,\mathrm{F_I'} \, e g_r G'\label{eq:dia2}\ ,
\end{eqnarray}
with
\begin{eqnarray}
b_2&=&\frac{i}{16\pi^2}\int^1_0 dx\int^x_0 dy \frac{y(y-2)}{s'}\nonumber\\
c_2&=&\frac{i}{16\pi^2}\int^1_0 dx \int^x_0 dy \frac{x(2-y)}{s'}\label{eq:bc}
\end{eqnarray}
and
\begin{eqnarray}
s'=-(P^0)^2 x^2+2P^0k^0 xy+((P^0)^2-M_2^2+M^2_1)x+(-2P^0k^0+M^2_2-M_l^2)y-M^2_1\label{eq:s1}\ .
\end{eqnarray}

The sum of the diagrams in Figs. \ref{fig:ppv} and \ref{fig:3v}, from Eqs. (\ref{eq:dia1}), (\ref{eq:c}), (\ref{eq:dia2}), (\ref{eq:bc}), gives rise to the following amplitude:
\begin{eqnarray}
-it_{K^{*+}(1430)\to K^+\gamma}^{ij}=\frac{1}{2}(b_2' k_\alpha P_\gamma+(c'-c_2')P_\alpha k_\gamma)(k^j\eps^{\alpha i\gamma\delta}+k^i\eps^{\alpha j\gamma\delta})\eps_\delta^{(\gamma)}eG'\ ,\label{eq:dia12}
\end{eqnarray}
with 
\begin{eqnarray}
b_2'&=&g_r\,\mathrm{F_I'} \,b_2\nonumber\\
c'&=&g_r\,\mathrm{F_I} \, c\nonumber\\
c_2'&=&g_r\,\mathrm{F_I'} \,c_2\ .
\label{eq:abc2}
\end{eqnarray}

In order to compute the decay width of the process $K^{*+}(1430)\to K^+\gamma$, we need to evaluate the squared amplitude summing over polarizations, i. e., $\frac{1}{2J+1}\sum_{\lambda_f}\sum_{\lambda_i}t_{ij} (t^{ij})^*$. The sum over the polarizations of the photon $\sum_{\lambda_f}\eps_\delta^{(\gamma)}\eps_{\delta'}^{(\gamma)}$ leads to a factor $-g_{\delta\delta'}$. In addition, products of the antisymmetric $\eps^{\alpha\beta\gamma\delta}$ operators appear, for what we make use of the rule
\begin{equation}
\renewcommand{\tabcolsep}{1cm}
\renewcommand{\arraystretch}{2}
\eps^{\alpha\beta\gamma\delta}\eps^{\alpha'\hspace{0.15cm}\gamma'\hspace{0.15cm}}_{\hspace{0.15cm}\beta'\hspace{0.15cm}\delta}=
-\left |
\begin{array}{ccc}
g^{\alpha\alpha'} & g^{\alpha}_{\hspace{0.15cm}\beta'}&g^{\alpha\gamma'}\\
g^{\beta\alpha'} & g^{\beta}_{\hspace{0.15cm}\beta'}&g^{\beta\gamma'}\\
g^{\gamma\alpha'} & g^{\gamma}_{\hspace{0.15cm}\beta'}&g^{\gamma\gamma'}
\end{array}
\right |\ ,
\end{equation}
with $\beta,\beta'$ spatial indices. Finally, we find
\begin{eqnarray}
\frac{1}{2J+1}\sum_{\lambda_f}\sum_{\lambda_i}\vert t\vert^2 =\frac{1}{2J+1}\vert \vec{k}\vert^4 P_0^2  | b'_2+c'_2-c'|^2(eG')^2 \ .
\end{eqnarray}
The $K^{*+}(1430)\to K^+\gamma$ decay width is 
\begin{eqnarray}
\Gamma(K^{*+}(1430)\to K^+\gamma)=\frac{1}{8\pi}\frac{1}{2J+1}\vert \vec{k}\vert^5 | b'_2+c'_2-c'|^2 (eG')^2\ .
\end{eqnarray}
We must include not only the $K^*\rho$ channel, but all the possible channels listed in Table \ref{tab:geng2}: the $K^*\omega$ and $K^*\phi$ channels. The different $\mathrm{F_I}$, $\mathrm{F_I'}$ for each channel $r$ are listed in Tables \ref{tab:1}, \ref{tab:2}, \ref{tab:3}, \ref{tab:4}, \ref{tab:5} and \ref{tab:6}. Therefore,
\begin{eqnarray}
c'&=&\frac{1}{16\pi^2} \int^1_0 dx\int^x_0 dy \, (1-x)(y-2)\sum_r \frac{\mathrm{F_I}(r)g_r}{s(r)}\nonumber\\
b_2'&=&\frac{1}{16\pi^2}\int^1_0 dx\int^x_0 dy\,  y(y-2)\sum_r \frac{\mathrm{F_I'}(r)g_r}{s'(r)}\nonumber\\
c_2'&=&\frac{1}{16\pi^2}\int^1_0 dx \int^x_0 dy\, x(2-y)\sum_r\frac{\mathrm{F_I'}(r)g_r}{s'(r)}\ .\label{eq:abcf}
\end{eqnarray}
For completeness, we show $s$, $s'$, $\mathrm{F_I}$ and $\mathrm{F_I'}$ again,
\begin{eqnarray}
& &s=-(P^0)^2 x^2+2P^0k^0 xy+((P^0)^2-M_2^2+M^2_1)x+(-2P^0k^0+M^2_2-m_l^2)y-M^2_1\nonumber\\
& &s'=-(P^0)^2 x^2+2P^0k^0 xy+((P^0)^2-M_2^2+M^2_1)x+(-2P^0k^0+M^2_2-M_l^2)y-M^2_1\nonumber\\
& &\mathrm{F_I}=\frac{1}{\sqrt{2}}AB\lambda g_I\nonumber\\
& &\mathrm{F_I'}=-\frac{1}{\sqrt{2}}g_I B D\lambda
\end{eqnarray}

\section{Results}

We evaluate the results for the two reactions where there are data, the 
$K^{*+}\to K^+\gamma$ and the $K^{*0}\to K^0\gamma$ decays. We include
the four  types of diagrams of Fig. \ref{fig:diag2}, where there is
exchange of pseudoscalar or vector mesons, with or without strangeness,
taking into account the three channels to which the resonance couples,
$\rho K^*$, $\omega K^*$ and $\phi K^*$. All the possibilities from
Fig. \ref{fig:diag} and partial widths for each different loop are given
in Tables \ref{tab:1}, \ref{tab:2}, \ref{tab:3}, \ref{tab:4},
\ref{tab:5} and \ref{tab:6}. The total sum of all the contributions of
these diagrams is mostly constructive for the $K^{*+}(1430)$. In
contrast, the interference between these diagrams is very destructive in
the case of the $K^{*0}(1430)$. We have evaluated the uncertainties in
the theoretical decay widths by assuming errors in the coupling
constants $\Delta g$ which were found to be of order $15\%$ in Ref.~\cite{geng}.
The errors in $\Gamma$ are obtained generating random
numbers of the couplings $g_i$ weighted by the Normal (Gaussian)
distribution: 
\begin{equation}
 f(x)=\frac{1}{\sigma\sqrt{2 \pi}} e^{\frac{(x-g)^2}{2\sigma^2}}\ ,
\end{equation}
for what the von Newmann rejection method is used. The average value of a sample of $30$ results and its standard deviation are taken for $\Gamma$ and its uncertainty.
The results that we get are: 
\begin{eqnarray}
\Gamma(K^{*+}\to K^+\gamma)&=&150\pm50\,\mathrm{KeV}\nonumber\\
\Gamma(K^{*0}\to K^0\gamma)&=&(1.0\pm0.8)\times 10^{-2}\,\mathrm{KeV}\ .\label{eq:th}
\end{eqnarray}
These results should be compared with the experimental widths 
\begin{eqnarray}
\Gamma(K^{*+}\to K^+\gamma)&=&236\pm50\,\mathrm{KeV}\nonumber\\
\Gamma(K^{*0}\to K^0\gamma)&=&<98 ~\,\mathrm{KeV}\ ,
\end{eqnarray}
where we have summed in quadrature the error in the branching ratio of
Eq. (\ref{eq:widex}) and the one of the total width of the PDG. As we
can see, the result for the charged $K^{*+}$ is compatible with the data
within errors and for the one of the neutral $K^{*0}$ the upper bound is
fulfilled. It would be interesting to have this upper bound improved 
experimentally, since we predict such a small number for the width.   
The contribution of the $PPV$ and $3V$ diagrams depicted in Fig. \ref{fig:diag} are
\begin{eqnarray}
 \Gamma_{(K^{*+}(1430)\to K^+\gamma), PPV}&=&46.6\,\mathrm{KeV}\nonumber\\
 \Gamma_{(K^{*+}(1430)\to K^+\gamma), 3V}&=&28.2\,\mathrm{KeV}\ ,\nonumber\\
 \Gamma_{(K^{*0}(1430)\to K^0\gamma), PPV}&=&0.19\,\mathrm{KeV}\nonumber\\
 \Gamma_{(K^{*0}(1430)\to K^0\gamma), 3V}&=&0.29\,\mathrm{KeV}\ .
\end{eqnarray}
It is interesting to see that the contributions of the two mechanisms are of the same order of magnitude, which means that both are important and must be taken into account to get the results of Eq. (\ref{eq:th}) within the range of the data in the PDG. One can observe, however, that in the case of the $K^{*+}$ the interference is constructive, while in the case of the $K^{*0}$ it is destructive.

\section{Conclusions}
 We have studied the decay of the $K_2^{*+}(1430)$ and $K_2^{*0}(1430)$
into a photon and a pseudoscalar meson. The states considered are 
those generated dynamically from the vector-vector
interaction in \cite{geng} that can be clearly assigned to known
resonances and that decay in this mode. 
The evaluation of the width
required the consideration of loop diagrams involving the coupling of
the resonances to the constituent vector-vector channels, plus some
anomalous couplings.
We find that the loops become convergent and we can evaluate finite values 
for the decay rates by making an approximation consistent with the VV molecular 
picture.  
The results obtained for the width of the $K_2^{*+}(1430)$ are well
within the experimental values considering errors. For the case of the
$K_2^{*0}(1430)$ we found a very small width, well below the
experimental upper bound of the PDG. It would be worth trying to
improve on this boundary since our results are so much smaller than the
present bound. The two results obtained
are adding progressive support to the
idea of the $K_2^{*}(1430)$ and other related resonances found in
\cite{geng} as dynamically generated from the vector-vector
interaction.


\begin{table}[H]
\begin{center}
\begin{tabular}{ccccc|ccccc|c}
$V_1$&$V_2$&$P_l$\T\B&$V_f$&$P_f$&A&B&$\lambda$&$g_I$&$\mathrm{F_I}$&$\Gamma_i(\mathrm{KeV})$\\\hline\hline
$K^{*0}$&$\rho^+$&$\pi^-$&\T\B$\omega$&$K^+$&$-1$&$\sqrt{2}$&$\frac{1}{3\sqrt{2}}$&$-\sqrt{\frac{2}{3}}$&$\frac{1}{3\sqrt{3}}$&$6.05$\\
$K^{*+}$&$\rho^0$&$\pi^0$&$\omega$&$K^+$&$-\frac{1}{\sqrt{2}}$\T\B&$\sqrt{2}$&$\frac{1}{3\sqrt{2}}$&$-\frac{1}{\sqrt{3}}$&$\frac{1}{6\sqrt{3}}$& \T\B\\\hline
$\rho^0$&$K^{*+}$&$K^-$&$\rho^0$&$K^+$&$\frac{1}{\sqrt{2}}$&\T\B$\frac{1}{\sqrt{2}}$&$\frac{1}{\sqrt{2}}$&$-\frac{1}{\sqrt{3}}$&$-\frac{1}{4\sqrt{3}}$\T\B&$5.05$\\
 & & &$\omega$& & &$\frac{1}{\sqrt{2}}$&$\frac{1}{3\sqrt{2}}$\T\B & &$-\frac{1}{12\sqrt{3}}$&\\
 & & &$\phi$& & &$1$&$-\frac{1}{3}$& &$\frac{1}{6\sqrt{3}}$\T\B&\\
$\rho^+$&$K^{*0}$&$\bar{K}^0$&$\rho^0$&$K^+$&$1$&\T\B$-\frac{1}{\sqrt{2}}$&$\frac{1}{\sqrt{2}}$&$-\sqrt{\frac{2}{3}}$&$\frac{1}{2\sqrt{3}}$\T\B&\\
& & &$\omega$& & &$\frac{1}{\sqrt{2}}$&$\frac{1}{3\sqrt{2}}$\T\B& &$-\frac{1}{6\sqrt{3}}$&\\
 & & &$\phi$& & &$1$&$-\frac{1}{3}$& &$\frac{1}{3\sqrt{3}}$\T\B&\\\hline
$K^{*+}$&$\rho^0$&$\eta$&$\rho^0$&$K^{+}$&$-\frac{2}{\sqrt{3}}$&$\frac{2}{\sqrt{3}}$&$\frac{1}{\sqrt{2}}$&$-\frac{1}{\sqrt{3}}$&$\frac{2}{3\sqrt{3}}$\T\B&$6.78$\\\hline
 & &$\eta'$& & &$\frac{1}{\sqrt{6}}$&$\sqrt{\frac{2}{3}}$&$\frac{1}{\sqrt{2}}$&$-\frac{1}{\sqrt{3}}$&$-\frac{1}{6\sqrt{3}}$\T\B&$0.24 $\\\hline\hline
\end{tabular}
\end{center}
\caption{$K^{*+}$ decay diagrams involving the $\rho K^*$ channel and the $PPV$ vertex.}
\label{tab:1}
\end{table}

\begin{table}[H]
\begin{center}
\begin{tabular}{ccccc|ccccc|c}
$V_1$&$V_2$&$P_l$\T\B&$V_f$&$P_f$&A&B&$\lambda$&$g_I$&$\mathrm{F_I}$&$\Gamma_i(\mathrm{KeV})$\\\hline\hline
$K^{*+}$&$\omega$&$\pi^0$&$\rho^0$&$K^+$&$-\frac{1}{\sqrt{2}}$&$\sqrt{2}$&$\frac{1}{\sqrt{2}}$&$1$&$-\frac{1}{2}$&$0.77$\T\B\\\hline
$\omega$&$K^{*+}$&$K^-$&$\rho^0$&$K^+$&$\frac{1}{\sqrt{2}}$&$\frac{1}{\sqrt{2}}$&$\frac{1}{\sqrt{2}}$&$1$&$\frac{1}{4}$\T\B&$0.07$\\
  & & &$\omega$& & &$\frac{1}{\sqrt{2}}$&$\frac{1}{3\sqrt{2}}$& &$\frac{1}{12}$\T\B&\\
  & & &$\phi$& & &$1$&$-\frac{1}{3}$& &$-\frac{1}{6}$\T\B&\\ \hline
$K^{*+}$&$\omega$&$\eta$&$\omega$&$K^+$&$-\frac{2}{\sqrt{3}}$&$\frac{2}{\sqrt{3}}$&$\frac{1}{3\sqrt{2}}$&$1$&$-\frac{2}{9}$\T\B&$9.5\times 10^{-2}$\\
  & & &$\phi$& & & & & &$0$\T\B\\\hline
 & &$\eta'$&$\omega$& &$\frac{1}{\sqrt{6}}$&$\sqrt{\frac{2}{3}}$&$\frac{1}{3\sqrt{2}}$&$1$&$\frac{1}{18}$\T\B&$3.2\times 10^{-3}$\\
  & & &$\phi$& & & & & &$0$\T\B&\\\hline
$\phi$&$K^{*+}$&$K^-$&$\rho^0$&$K^+$&$-1$&$\frac{1}{\sqrt{2}}$&$\frac{1}{\sqrt{2}}$&$1$&$-\frac{1}{2\sqrt{2}}$&$8.4\times 10^{-2}$\T\B\\
  & & &$\omega$& & &$\frac{1}{\sqrt{2}}$&$\frac{1}{3\sqrt{2}}$& &$-\frac{1}{6\sqrt{2}}$&\T\B\\
 & & &$\phi$& & &$1$&$-\frac{1}{3}$& &$\frac{1}{3\sqrt{2}}$\T\B&\\\hline
$K^{*+}$&$\phi$&$\eta$&$\omega$&$K^+$& & & & &$0$&$0.17$\T\B\\
  & & &$\phi$& &$-\frac{2}{\sqrt{3}}$&$-\frac{2}{\sqrt{3}}$&$-\frac{1}{3}$&$1$&$-\frac{2\sqrt{2}}{9}$&\T\B\\\hline
 & &$\eta'$&$\omega$& & & & & &$0$\T\B&$2.6\times 10^{-2}$\\
  & & &$\phi$& &$\frac{1}{\sqrt{6}}$&$2\sqrt{\frac{2}{3}}$&$-\frac{1}{3}$&$1$&$-\frac{\sqrt{2}}{9}$\T\B&\\\hline\hline
\end{tabular}
\end{center}
\caption{$K^{*+}$ decay diagrams involving the $\omega K^*$ and $\phi K^{*}$ channels and the $PPV$ vertex.}
\label{tab:2}
\end{table}

\begin{table}[H]
\begin{center}
\begin{tabular}{ccccc|cccccc}
$V_1$&$V_2$&$P_l$\T\B&$V_f$&$P_f$&A&B&$\lambda$&$g_I$&$\mathrm{F_I}$&$\Gamma_i(\mathrm{KeV})$\\\hline\hline
$K^{*0}$&$\rho^0$&$\pi^0$&\T\B$\omega$&$K^0$&$\frac{1}{\sqrt{2}}$&$\sqrt{2}$&$\frac{1}{3\sqrt{2}}$&$\frac{1}{\sqrt{3}}$&$\frac{1}{6\sqrt{3}}$&$6.05$\\
$K^{*+}$&$\rho^-$&$\pi^+$&$\omega$&$K^0$&$-1$\T\B&$\sqrt{2}$&$\frac{1}{3\sqrt{2}}$&$-\sqrt{\frac{2}{3}}$&$\frac{1}{3\sqrt{3}}$\T\B&\\\hline
$\rho^-$&$K^{*+}$&$K^-$&$\rho^0$&$K^0$&$1$&\T\B$\frac{1}{\sqrt{2}}$&$\frac{1}{\sqrt{2}}$&$-\sqrt{\frac{2}{3}}$&$-\frac{1}{2\sqrt{3}}$\T\B&$0$\\
 & & &$\omega$& & &$\frac{1}{\sqrt{2}}$&$\frac{1}{3\sqrt{2}}$\T\B& &$-\frac{1}{6\sqrt{3}}$&\\
 & & &$\phi$& & &$1$&$-\frac{1}{3}$& &$\frac{1}{3\sqrt{3}}$\T\B&\\
$\rho^0$&$K^{*0}$&$\bar{K}^0$&$\rho^0$&$K^0$&$-\frac{1}{\sqrt{2}}$&\T\B$-\frac{1}{\sqrt{2}}$&$\frac{1}{\sqrt{2}}$&$\frac{1}{\sqrt{3}}$&$\frac{1}{4\sqrt{3}}$&\T\B\\
 & & &$\omega$& & &$\frac{1}{\sqrt{2}}$&$\frac{1}{3\sqrt{2}}$\T\B & &$-\frac{1}{12\sqrt{3}}$&\\
 & & &$\phi$& & &$1$&$-\frac{1}{3}$& &$\frac{1}{6\sqrt{3}}$\T\B&\\\hline
$K^{*0}$&$\rho^0$&$\eta$&$\rho^0$&$K^{0}$&$-\frac{2}{\sqrt{3}}$&$\frac{2}{\sqrt{3}}$&$\frac{1}{\sqrt{2}}$&$\frac{1}{\sqrt{3}}$&$-\frac{2}{3\sqrt{3}}$\T\B&$6.78$\\
 & &$\eta'$& & &$\frac{1}{\sqrt{6}}$&$\sqrt{\frac{2}{3}}$&$\frac{1}{\sqrt{2}}$&$\frac{1}{\sqrt{3}}$&$\frac{1}{6\sqrt{3}}$&$0.24$\T\B\\\hline\hline
\end{tabular}
\end{center}
\caption{$K^{*0}$ decay diagrams involving the $\rho K^*$ channel and the $PPV$ vertex.}
\label{tab:3}
\end{table}

\begin{table}[H]
\begin{center}
\begin{tabular}{ccccc|cccccc}
$V_1$&$V_2$&$P_l$\T\B&$V_f$&$P_f$&A&B&$\lambda$&$g_I$&$\mathrm{F_I}$&$\Gamma_i(\mathrm{KeV})$\\\hline\hline
$K^{*0}$&$\omega$&$\pi^0$&$\rho^0$&$K^0$&$\frac{1}{\sqrt{2}}$&$\sqrt{2}$&$\frac{1}{\sqrt{2}}$&$1$&$\frac{1}{2}$&$0.77$\T\B\\\hline
$\omega$&$K^{*0}$&$\bar{K}^0$&$\rho^0$&$K^0$&$\frac{1}{\sqrt{2}}$&$-\frac{1}{\sqrt{2}}$&$\frac{1}{\sqrt{2}}$&$1$&$-\frac{1}{4}$&$0.28$\T\B\\
 & & &$\omega$& & &$\frac{1}{\sqrt{2}}$&$\frac{1}{3\sqrt{2}}$& &$\frac{1}{12}$&\T\B\\
  & & &$\phi$& & &$1$&$-\frac{1}{3}$& &$-\frac{1}{6}$&\T\B\\ \hline
$K^{*0}$&$\omega$&$\eta$&$\omega$&$K^0$&$-\frac{2}{\sqrt{3}}$&$\frac{2}{\sqrt{3}}$&$\frac{1}{3\sqrt{2}}$&$1$&$-\frac{2}{9}$\T\B&$9.5\times 10^{-2}$\\
  & & &$\phi$& & & & & &$0$\T\B&\\\hline
& &$\eta'$&$\omega$& &$\frac{1}{\sqrt{6}}$&$\sqrt{\frac{2}{3}}$&$\frac{1}{3\sqrt{2}}$&$1$&$\frac{1}{18}$\T\B&$3.4\times 10^{-3}$\\
 & & &$\phi$& & & & & &$0$&\T\B\\\hline
$\phi$&$K^{*0}$&$\bar{K}^0$&$\rho^0$&$K^0$&$-1$&$-\frac{1}{\sqrt{2}}$&$\frac{1}{\sqrt{2}}$&$1$&$\frac{1}{2\sqrt{2}}$\T\B&$0.34$\\
  & & &$\omega$& & &$\frac{1}{\sqrt{2}}$&$\frac{1}{3\sqrt{2}}$& &$-\frac{1}{6\sqrt{2}}$\T\B&\\
  & & &$\phi$& & &$1$&$-\frac{1}{3}$& &$\frac{1}{3\sqrt{2}}$\T\B&\\\hline
$K^{*0}$&$\phi$&$\eta$&$\omega$&$K^0$& & & & &$0$\T\B&$0.17$\\
  & & &$\phi$& &$-\frac{2}{\sqrt{3}}$&$-\frac{2}{\sqrt{3}}$&$-\frac{1}{3}$&$1$&$-\frac{2\sqrt{2}}{9}$\T\B&\\\hline
& &$\eta'$&$\omega$& & & & & &$0$\T\B&$2.6\times 10^{-2}$\\
  & & &$\phi$& &$\frac{1}{\sqrt{6}}$&$2\sqrt{\frac{2}{3}}$&$-\frac{1}{3}$&$1$&$-\frac{\sqrt{2}}{9}$&\T\B\\\hline\hline
\end{tabular}
\end{center}
\caption{$K^{*0}$ decay diagrams involving the $\omega K^*$ and $\phi K^{*}$ channels and the $PPV$ vertex.}
\label{tab:4}
\end{table}


\begin{table}[H]
\begin{center}
\begin{tabular}{ccccc|ccccc|c}
$V_1$&$V_2$&$V_f$\T\B&$V_l$&$P_f$&D&B&$\lambda$&$g_I$&$\mathrm{F_I'}$&$\Gamma_i(\mathrm{KeV})$\\\hline\hline
$K^{*0}$&$\rho^+$&$\rho^0$&\T\B$\rho^-$&$K^+$&$\sqrt{2}$&$1$&$\frac{1}{\sqrt{2}}$&$-\sqrt{\frac{2}{3}}$&$\frac{1}{\sqrt{3}}$&$12.8$\\\hline
$\rho^0$&$K^{*+}$&$\rho^0$&$K^{*-}$&$K^+$&$\frac{1}{\sqrt{2}}$\T\B&$\frac{1}{\sqrt{2}}$&$\frac{1}{\sqrt{2}}$&$-\frac{1}{\sqrt{3}}$&$\frac{1}{4\sqrt{3}}$&$2.31$\T\B\\
  & &$\omega$& & &$\frac{1}{\sqrt{2}}$&\T\B &$\frac{1}{3\sqrt{2}}$& &$\frac{1}{12\sqrt{3}}$&\T\B\\
  & &$\phi$& & &$-1$& &$-\frac{1}{3}$\T\B & &$\frac{1}{6\sqrt{3}}$&\\\hline
%
\T\B$\omega$&$K^{*+}$&$\rho^0$&$K^{*-}$&$K^+$&$\frac{1}{\sqrt{2}}$&$\frac{1}{\sqrt{2}}$&$\frac{1}{\sqrt{2}}$&$1$&$-\frac{1}{4}$&$0.29$\\
  & &$\omega$& & &\T\B$\frac{1}{\sqrt{2}}$& &$\frac{1}{3\sqrt{2}}$& &$-\frac{1}{12}$&\\
  & &$\phi$& & &\T\B$-1$& &$-\frac{1}{3}$& &$-\frac{1}{6}$&\\\hline
\T\B$\phi$&$K^{*+}$&$\rho^0$&$K^{*-}$&$K^+$&$\frac{1}{\sqrt{2}}$&$1$&$\frac{1}{\sqrt{2}}$&$1$&$-\frac{1}{2\sqrt{2}}$&$0.56$\\
  & &$\omega$& & &\T\B$\frac{1}{\sqrt{2}}$& &$\frac{1}{3\sqrt{2}}$& &$-\frac{1}{6\sqrt{2}}$&\\
 & &$\phi$& & &\T\B$-1$& &$-\frac{1}{3}$& &$-\frac{1}{3\sqrt{2}}$&\\
\hline\hline
\end{tabular}
\end{center}
\caption{$K^{*+}$ decay diagrams involving the $3V$ vertex.
Terms which involve a $\gamma K^*K^*$ coupling with a neutral $K^*$ are
 zero and are omitted from the table.
}
\label{tab:5}
\end{table}

\begin{table}[H]
\begin{center}
\begin{tabular}{ccccc|cccccc}
$V_1$&$V_2$&$V_f$\T\B&$V_l$&$P_f$&D&B&$\lambda$&$g_I$&$\mathrm{F_I'}$&$\Gamma_i(\mathrm{KeV})$\\\hline\hline
$K^{*+}$&$\rho^-$&$\rho^0$&\T\B$\rho^+$&$K^0$&$-\sqrt{2}$&$1$&$\frac{1}{\sqrt{2}}$&$-\sqrt{\frac{2}{3}}$&$-\frac{1}{\sqrt{3}}$&$12.8$\\\hline
$\rho^-$&$K^{*+}$&$\rho^0$&$K^{*-}$&$K^0$&$\frac{1}{\sqrt{2}}$&\T\B$1$&$\frac{1}{\sqrt{2}}$&$-\sqrt{\frac{2}{3}}$&$\frac{1}{2\sqrt{3}}$&9.27\\
 & &$\omega$& & &$\frac{1}{\sqrt{2}}$& &\T\B$\frac{1}{3\sqrt{2}}$& &$\frac{1}{6\sqrt{3}}$&\\
& &$\phi$& & &$-1$& &$-\frac{1}{3}$& &$\frac{1}{3\sqrt{3}}$&\T\B\\\hline
\hline\hline
\end{tabular}
\end{center}
\caption{$K^{*0}$ decay diagrams involving the $3V$ vertex. Terms which involve a $\gamma K^*K^*$ coupling with a neutral $K^*$ are
 zero and are omitted from the table.}
\label{tab:6}
\end{table}

\end{document}